\documentclass[pss]{wiley2sp} 
\usepackage{amsmath}

\begin{document}

\title{Pentagonal puckering in a sheet of amorphous graphene}

\titlerunning{Electronic properties of amorphous graphene }

\author{%
  Y. Li\textsuperscript{\textsf{\bfseries 1}},
  F. Inam\textsuperscript{\textsf{\bfseries 2}},
  A. Kumar\textsuperscript{\textsf{\bfseries 3}},
  M. F. Thorpe\textsuperscript{\textsf{\bfseries 3}},
  D. A. Drabold\textsuperscript{\Ast,\textsf{\bfseries 1}}}

\authorrunning{Y. Li et al.}

\mail{e-mail
  \textsf{drabold@ohio.edu}, Phone: +1 740 593 1715, Fax: +1 740 593 0433}

\institute{%
  \textsuperscript{1}\, Department of Physics and Astronomy, Ohio University, Athens, OH 45701, USA\\
  \textsuperscript{2}\, The Abdus Salam ICTP, Strada Costiera 11, Trieste 34151, ITALY\\
  \textsuperscript{3}\, Department of Physics and Astronomy, Arizona State University, Tempe, AZ 85287, USA\\}

\received{XXXX, revised XXXX, accepted XXXX} 
\published{XXXX} 

\abstract{

Ordered graphene has been extensively studied. In this paper we undertake a first density functional study of {\it topologically disordered} analogues of graphene, in the form of a random network, consisting predominantly of hexagonal rings, but also including pentagons and heptagons. After some preliminaries with crystalline material, we relax various random network models and find that the presence of carbon pentagons induce local curvature, thus breaking the initial planar symmetry, in some analogy with the case of fullerenes. Using density functional theory to calculate the total energy, we find that while the planar state is locally stable, there is a puckered state that has lower energy. The scale of the puckering is consistent with that expected with local maxima and minima associated with pentagons surrounded by larger rings; forming local ''buckyball domes''.}

\maketitle  

\section{Introduction}
Graphene is among the hottest topics in current condensed matter science. A vast amount of work on many aspects of crystalline graphene has appeared. In this paper we take a different tack: we explore the role of topological disorder in amorphous graphene. 

The structure of conventional amorphous semiconductors like amorphous Si or Ge is well represented
by the continuous random network model (CRN) introduced
by Zachariasen\cite{JACS54} 70 years ago. The CRN model has the
simplicity that each of the atoms should satisfy its local
bonding requirements, and should have minimum strain, characterized by having
a narrow bond angle and bond-length distribution. Recently there has been an amorphous graphene CRN model proposed \cite{PSSB247}. Here we develop Kapko et al.'s work \cite{PSSB247}, and show that pentagons induce curvature in the free standing sheets and analyze the electronic properties.

\section{Models}

To calculate the band structure, the primitive cell of graphene was used. To calculate the density of states of crystalline graphene, a 800-atom model (800 c-g) was made. For amorphous graphene, we used one 800-atom (800 a-g) model and two 836-atom models (836 a-g1 and 836 a-g2). These amorphous graphene models were all prepared by a modified Wooten-Weaire-Winer (WWW) method\cite{PSSB247}.

\section{Crystalline graphene}

\subsection{Band structure}
When calculating the band structure of graphene, tight-binding and $ab$ $initio$ methods are two widely used tools. Reich et al. have compared the result of tight binding with $ab$ $initio$\cite{PRB66}. Nowadays, two of the widely used $ab$ $initio$ programs are SIESTA\cite{siesta}, using pseudopotentials and the Perdew-Zunger parameterization of the local-density approximation (LDA), and VASP\cite{vasp}, with pseudopotentials, plane-wave basis and LDA.

\begin{figure}[htb]%
\includegraphics*[width=\linewidth,height=.75\linewidth]{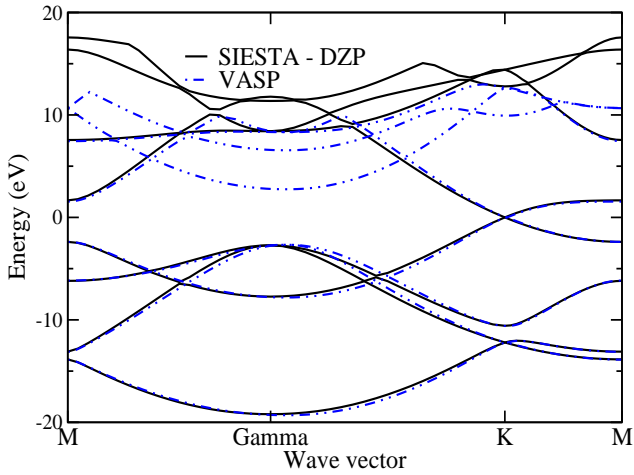}
\caption{%
  Band structure of graphene. a) In SIESTA, the solid line represents the result of DZP basis. b) The result of VASP is given by dotted-dashed line.}
\label{fig:band}
\end{figure}

We computed the eight lowest-energy bands of graphene, by using a single-$\zeta$ (SZ) basis set and a double-$\zeta$, polarized (DZP) basis set with SIESTA, and also VASP. 20 kpoints along each special symmetry line were taken for both SZ and DZP calculations by SIESTA, and 50 kpoints along each line for VASP. The results of SZ and DZP are essentially identical for occupied bands, and exhibit differences for the unoccupied states. Figure \ref{fig:band} shows SIESTA results using DZP basis and VASP. We compared our results with the other first principle calculations \cite{PRB75} \cite{PRB78}. The VASP and SIESTA results are in good agreement with published results for each code \cite{PRB75} \cite{PRB78}. However, as shown in Figure \ref{fig:band}, the calculation based on DZP is in good agreement with plane-wave pseudopotential calculation (VASP) for the four occupied low-energy bands, unlike the unoccupied higher energy range, where the agreement is rather poor. This discrepancy in the unoccupied bands is interesting, and we will explore this further elsewhere.

Computationally speaking, VASP is more time-consuming than SIESTA, particularly for large amorphous graphene models we discuss later. For computing total energies and forces, e.g. utilizing quantities soley from the occupied electronic subspace, SIESTA in SZ approximation is a reasonable choice.

\section{Amorphous graphene}

The three amorphous graphene models are prepared by introducing Stone-Wales defects into perfect honey comb lattice. The configuration of 800 a-g model is presented in \cite{PSSB247}, each atom is three-fold.

\begin{figure}[htb]%
\includegraphics*[width=\linewidth,height=.75\linewidth]{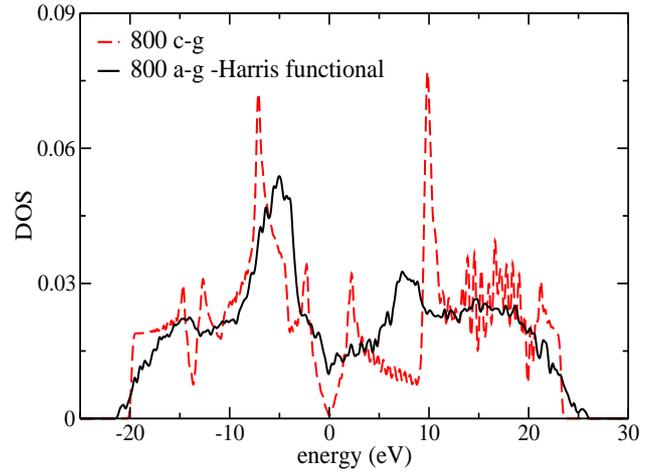}
\caption{%
Density of states of 800-atom amorphous and crystalline graphene, the Fermi energy is $0 eV$. a) Dot-dash line shows the result of 800 a-g model using Harris functional approximation by SIESTA. c) The density of states of 800 c-g model is shown by the dashed line, at the $\Gamma$ point of Brillouin zone (BZ). The ringing in the crystal is due to incomplete BZ sampling.}
\label{fig:compare_dos}
\end{figure}

\subsection{Density of states}

The electronic density of states for the initial flat 800 a-g model is compared to a $\Gamma$ point density of states for the crystalline 800 c-g model in Figure \ref{fig:compare_dos}, using SIESTA with a SZ basis and the Harris functional. From this figure, we observe that the electronic structure of 800 a-g model is vastly different from the crystalline graphene near the Fermi level, as first reported earlier by Kapko et al.\cite{PSSB247}. We have constructed additional models with periodic boundary conditions and 836 atoms each (836 a-g1 and 836 a-g2 models). The ring statistics of these three models are given in Table \ref{tab:ring} and these show some small differences.

\begin{table}[htb]
\centering
  \caption{Ring statistics of 800 a-g, 836 a-g1 and 836 a-g2 models, shown as \%.}
  \begin{tabular}[htbp]{@{}llll@{}}
    \hline
      Ring Size  & 800 a-g   &  836 a-g1  &   836 a-g2  \\
    \hline   
    5  &  33.5   &  25  &   24 \\
    6  &  38    &    53   &  52 \\
    7  &  24    &   19   &  25 \\
    8  &  4.5   &  3  &  0  \\
    \hline
  \end{tabular}
  \label{tab:ring}
\end{table}

\subsection{Loss of planar symmetry}

In all three amorphous graphene models, we introduced small random fluctuations in the coordinates, in the direction normal to the graphene plane,  and then relaxed with the Harris functional and a SZ basis set. Starting with a flat sheet, the planar symmetry breaks with curvature above or below initial the plane. The final distortion depends on the initial conditions. However, a consistent theme emerges of pentagons inducing curvature as we describe below.

\begin{figure}[htb]%
\includegraphics*[width=\linewidth,height=.75\linewidth]{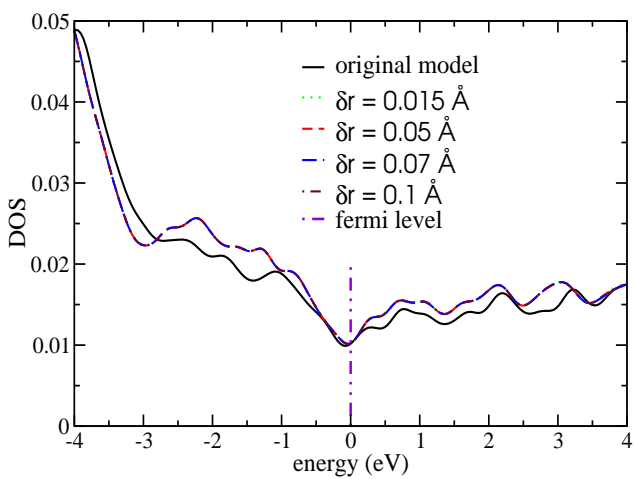}
\caption{%
Density of states of the original and relaxed crinkled system. a) The solid line is the density of states of original 800-atom amorphous graphene model. b) The density of states of crinkled systems are shown as marked in the plot. c) The Fermi level is corrected to $0 eV$ in the plot, as shown in dot-slash line.}
\label{fig:800a_crin_dos}
\end{figure}

\begin{figure}[htb]%
\includegraphics*[width=\linewidth,height=\linewidth]{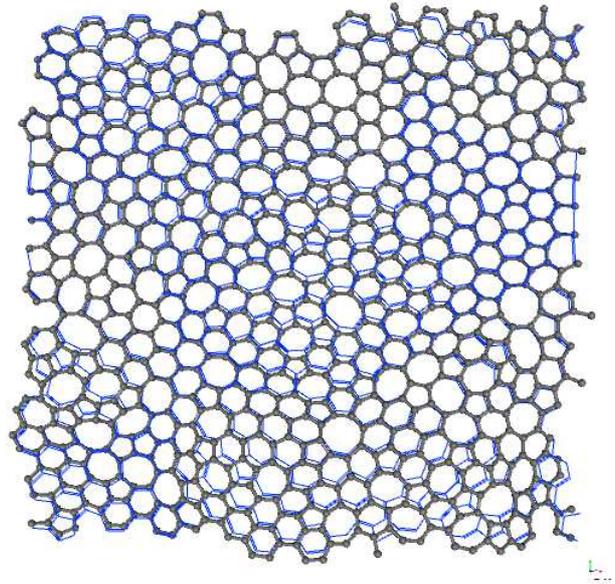}
\caption{%
The flat view of the relaxed 836 a-g1 system (in gray). The blue background illustrates the original 836 a-g1 model.}
\label{fig:crin_plot}
\end{figure}

\begin{figure}[htb]%
\includegraphics*[width=\linewidth,height=0.4\linewidth]{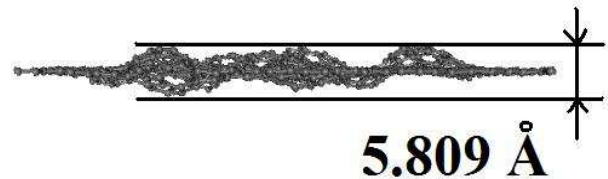}
\caption{%
The side view of the relaxed 800 a-g system. The biggest separation along normal direction is marked in the plot.}
\label{fig:crin_side}
\end{figure}

As shown in Table \ref{tab:table_Fermi_r}, Table \ref{tab:sample1_tab} and Table \ref{tab:sample2_tab}, first we randomly moved the atoms along normal direction in the range of $[-\delta r, +\delta r]$, as shown in the first column of these tables; and the results of relaxing by SIESTA in SZ basis are shown in the second, third and fourth columns. The prime symbol refers to the relaxed model. Taking the 800 a-g model as an example, the influence of puckering the system on the density of states around Fermi level is shown in Figure \ref{fig:800a_crin_dos}; an intuitive view of the fluctuation after relaxing is shown in Figure \ref{fig:crin_plot} and \ref{fig:crin_side} when $\delta r = 0.05\AA$. After breaking the planar symmetry by a tiny amount, say $\delta r = 0.05 \AA$, all three models pucker and form the rippled or undulated structure as shown in Figure \ref{fig:crin_plot} and \ref{fig:crin_side}.

\begin{table}[htb]
\centering
  \caption{the influence of $\delta$r on 800 a-g system relative to initial flat model}
  \begin{tabular}[htbp]{@{}lll@{}}
    \hline
    $\delta r$ ($\AA$) &  $\overline{\delta r'}$  ($\AA$) & $E_{tot}/N_{atom}$ ($eV$) \\
    \hline
    0.01   &  0.520  &   -0.107 \\
    0.05    &    0.525   &    -0.107 \\
    0.07    &   0.526   &   -0.107 \\
    \hline
  \end{tabular}
  \label{tab:table_Fermi_r}
\end{table}

\begin{table}[htb]
\centering
  \caption{the influence of $\delta$r on 836 a-g1 system relative to initial flat model}
  \begin{tabular}[htbp]{@{}lll@{}}
    \hline
    $\delta r$ ($\AA$) &  $\overline{\delta r'}$ ($\AA$) & $E_{tot}/N_{atom}$ ($eV$) \\
    \hline
    0.01     &   2.53E-3  &    0.0 \\
    0.05     &    1.402   &    -0.102 \\
    0.07     &   1.401  &    -0.102 \\
    \hline
  \end{tabular}
  \label{tab:sample1_tab}
\end{table}

\begin{table}[htb]
\centering
  \caption{the influence of $\delta$r on 836 a-g2 system relative to initial flat model}
  \begin{tabular}[htbp]{@{}lll@{}}
    \hline
    $\delta r$ ($\AA$) &   $\overline{\delta r'}$ ($\AA$) &  $E_{tot}/N_{atom}$ ($eV$) \\
    \hline
    0.01   &  2.72E-3  &  0.0 \\
    0.05   &    1.183   &   -0.090 \\
    0.07   &    1.180  &    -0.090 \\
    \hline
  \end{tabular}
  \label{tab:sample2_tab}
\end{table}

\begin{figure}[htb]%
\includegraphics*[width=\linewidth,height=.75\linewidth]{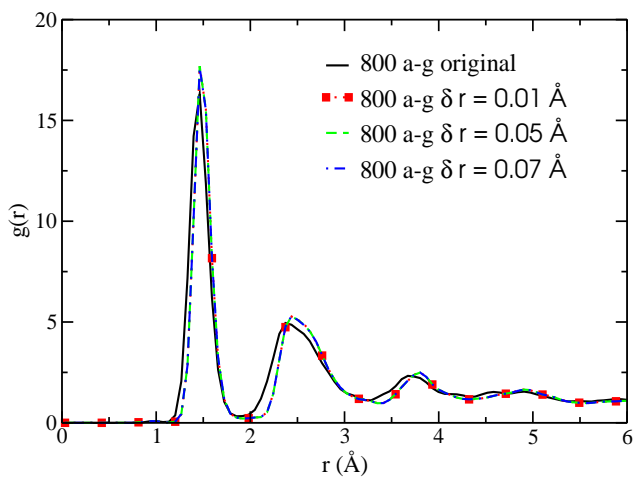}
\caption{%
Radial distribution function of flat and crinkled 800 a-g system.}
\label{fig:800a_crin_g(r)}
\end{figure}

The radial distribution function $g(r)$ is shown in Figure \ref{fig:800a_crin_g(r)}. From this plot, the mean bond length of the relaxed systems with different initial $\delta r$ remain near $1.42\AA$, the change in ring statistics after relaxing is also not significant. And according to Fig \ref{fig:crin_plot}, only one bond broke after relaxation. This implies that the main difference between the original amorphous graphene model and the relaxed ones is due to these undulations in Figure \ref{fig:crin_plot}. 

To compare with the puckering of the 800 a-g model, we also introduced the same planar symmetry breaking into a 800 c-g model and relaxed it. As expected, the atoms in this crystalline system maintained planar symmetry.

\begin{figure}[htb]%
\includegraphics*[width=\linewidth,height=.75\linewidth]{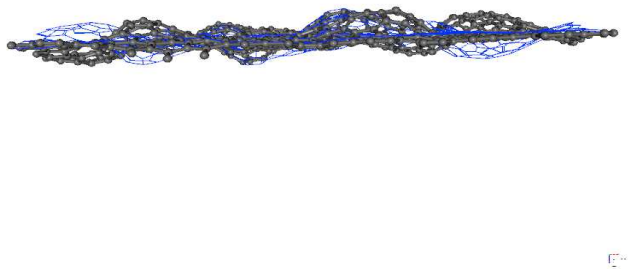}
\caption{%
The side view of the final configuration by using new and original RNG. a) The gray balls and sticks show the result of new RNG. b) The blue frames represent the result of original RNG.}
\label{fig:gen_side_0.05}
\end{figure}

Also in order to find the relation between the ripples in the relaxed systems and the initial random distortion, we tested different seeds in random number generator (RNG) and also different RNG. The results reveals that the changing seeds or or employing different RNG are quantitatively small: The maximum mean distortion from the original flat plane ($\overline{\delta r'}$) is about $0.545\AA$ and the maximum change in total energy is around $0.01 eV$ per atom. Figure \ref{fig:gen_side_0.05} shows the side view of the final configurations by using new and original RNG when $\delta r=0.05 \AA$, we can tell that the rippled regions are similar, except certain regions have formed "bucky domes" on opposite sides of the initial plane.

\begin{figure}[htb]%
\includegraphics*[width=\linewidth,height=.75\linewidth]{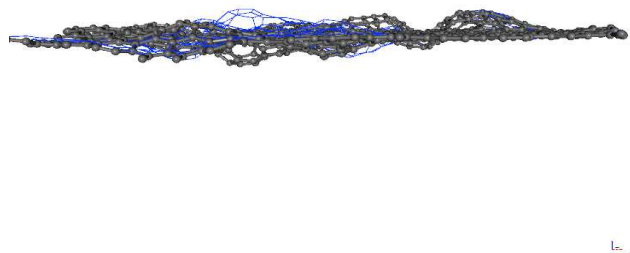}
\caption{%
The side view of the final configuration with $\delta r=0.05\AA$ of only moved atoms within pentagons and the original relaxation (distort all atoms). a) The gray balls and sticks show the result of moving atoms within pentagons. b) The blue frames represent the result of original distortion.}
\label{fig:ring_side_0.05}
\end{figure}

To further test the relation between the ripples and the initial distortion, instead of randomly moving all the atoms of 800 a-g, we only distorted the atoms within pentagons and compared with  atoms not included in pentagons. Figure \ref{fig:ring_side_0.05} shows the side view of the final configuration of relaxed 800 a-g system when the $\delta r = 0.01 \AA$ and only the atoms within pentagons were randomly moved . These results are similar as the previous test: a) The maximum change in $\overline{\delta r'}$ is around $0.625 \AA$ and the maximum change in total energy is around $0.02 eV$ per atom. b) No matter which atoms were distorted initially, the final puckered regions involve the same atoms, but possibly puckered in the opposite direction relative to the symmetry plane. Finally, we note that a 128-atom amorphous graphene model made with ''melt quenching'' \cite{EPJB681} exhibits regions puckered around pentagons in a similar fashion to what we report here.

\begin{figure}[ht]%
\includegraphics*[width=\linewidth,height=\linewidth]{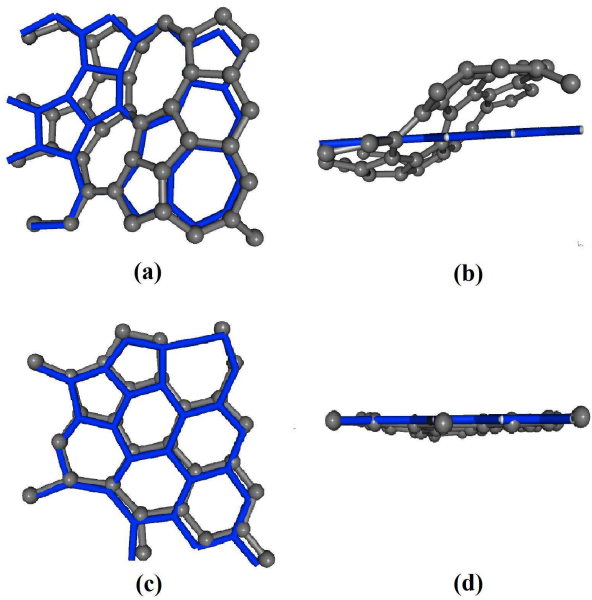}
\caption{%
The enlarged plot of crinkled and smooth region of 800 a-g model. a) The top view of the crinkled region. b) The side view of the crinkled region. c) The top view of the smooth region. d) The side view of the smooth region.}
\label{fig:800a-g_region}
\end{figure}

\begin{figure}[ht]%
\includegraphics*[width=\linewidth,height=\linewidth]{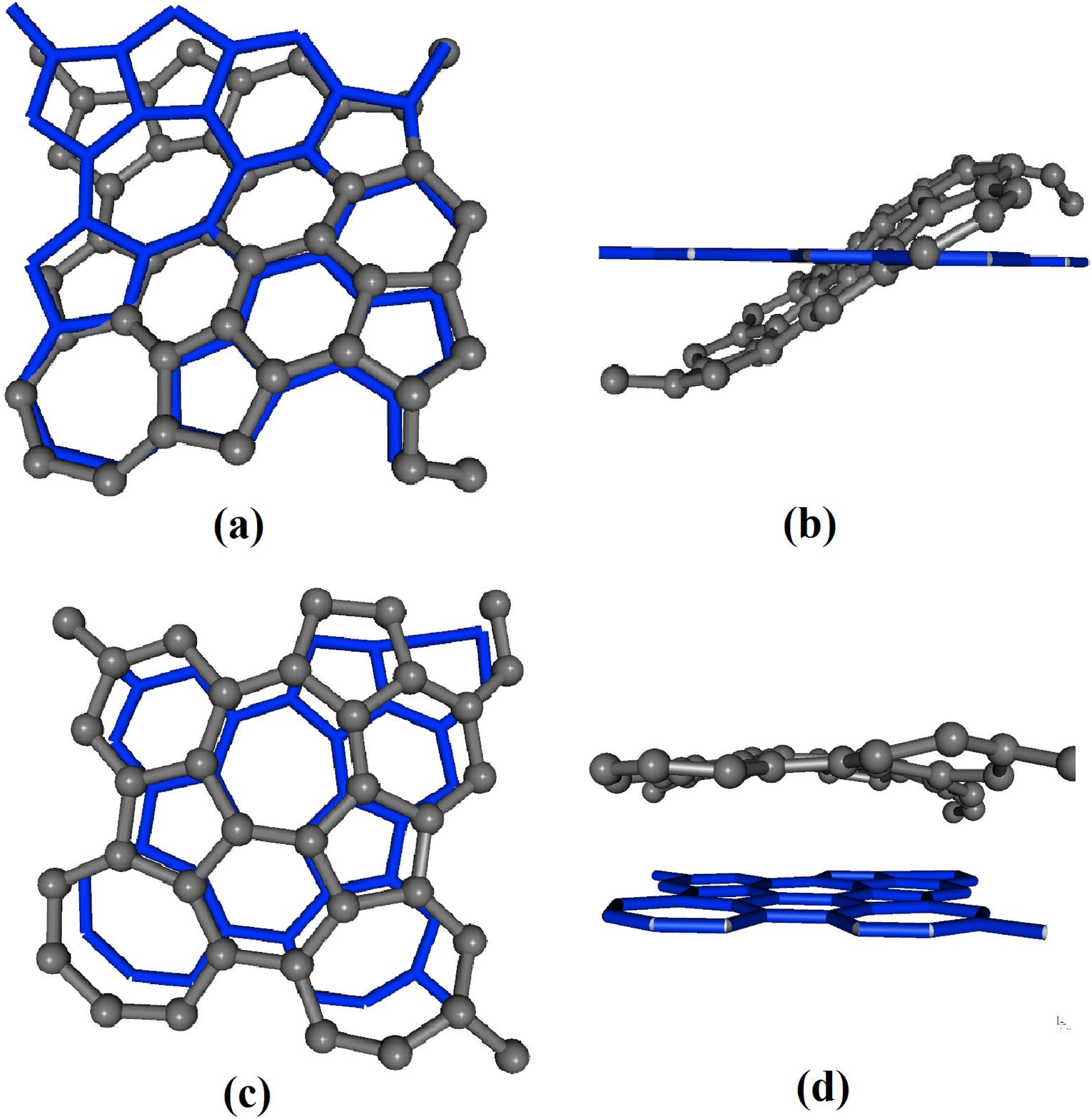}
\caption{%
The enlarged plot of crinkled and smooth region of 836 a-g1 model. a) The top view of the crinkled region. b) The side view of the crinkled region. c) The top view of the smooth region. d) The side view of the smooth region.}
\label{fig:sample1_region}
\end{figure}

\begin{figure}[ht]%
\includegraphics*[width=\linewidth,height=\linewidth]{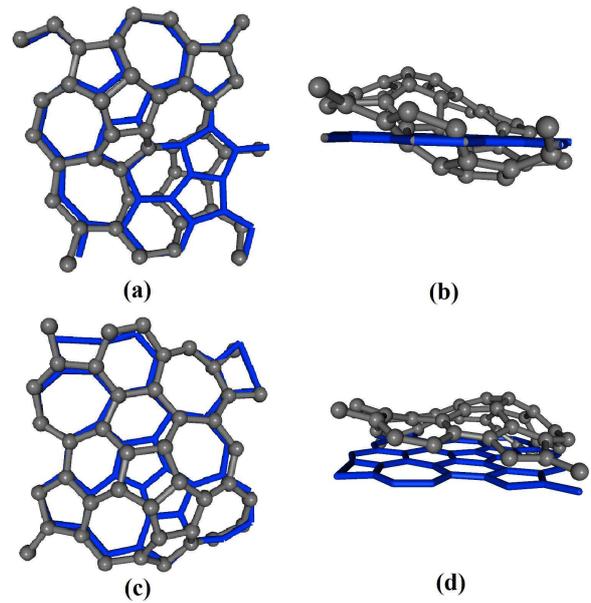}
\caption{%
The enlarged plot of crinkled and smooth region of 836 a-g2 model. a) The top view of the crinkled region. b) The side view of the crinkled region. c) The top view of the smooth region. d) The side view of the smooth region.}
\label{fig:sample2_region}
\end{figure}

Different initial symmetry breaking leads to different nearly degenerate states after relaxing. However, as stated above, no matter how different initial condition is ( or how different these degenerated state is), the puckered regions are almost the same. It is evident that different rings induce these ripples. With this motivation, we searched for regions where the height differences of two neighbor atoms are the largest and smallest in the model (crinkled and smooth regions), as shown in Figure \ref{fig:800a-g_region}, \ref{fig:sample1_region} and \ref{fig:sample2_region}. In these plots, the gray atoms are the configuration of crinkled system, and the blue straight lines represent the original model.

As illustrated in Figure \ref{fig:800a-g_region} and  \ref{fig:sample1_region}, the puckered areas are pentagon-dense areas. The bonds with most distortion do not belong to these pentagons, instead they are within the hexagons or heptagons connecting two pentagons. And the smooth areas have fewer pentagons than the crinkled areas, and most parts of the smooth areas contain hexagons and pentagons. These plots imply that the hexagons and heptagons alone will not lead to planar symmetry breaking. These ripples formed by pentagons strongly remind us of the fullerenes, especially the buckyball ($C_{60}$) which only contains pentagons and hexagons. The distance from the top to the bottom of the ripples for 800 a-g is around $5.809 \AA$ as shown in Figure \ref{fig:crin_side}, which is comparable to the diameter of buckyball, $6.636 \AA$. As shown in Figure \ref{fig:800a-g_region}, \ref{fig:sample1_region} and \ref{fig:sample2_region}, the crinkled regions are all associated with pentagons. 

\section{Conclusion}
We have shown that random network models of amorphous graphene pucker. While the planar conformation is locally stable, a lower energy solution is obtained that is puckered with local maxima and minima in the vicinity of pentagons. The relaxation is performed using a density functional calculation of the electronic energy. The scale of the puckering is consistent with the curvature found in buckyball caps with a pentagon surrounded by larger rings.

\begin{acknowledgement}
We want to thank  Dr. Mingliang Zhang, Bin Cai and Binay Prasai at Ohio University and Dr. Vitaliy Kapko at Arizona State University. All the calculations of radial distribution function, scattering structure factor and ring statistics are done by ISAACS\cite{petkov}. This work is supported by NSF Grant No. DMR 09-03225 and DMR 07-03973.

\end{acknowledgement}

\end{document}